 \documentclass[final,5p,times,twocolumn]{elsarticle}

\usepackage{amssymb}
\usepackage{amsthm}
\usepackage{amsmath}
\usepackage{eurosym}

\journal{Nuclear Physics A}

\begin{document}

\begin{frontmatter}



\title{A muon source based on plasma accelerators}



\author[infn]{L.~Serafini}
\author[infn]{I.~Drebot\corref{cor1}}
\ead{illya.drebot@mi.infn.it}
\cortext[cor1]{Corresponding author}

\author[infn]{A.~Bacci\corref{}}
\author[infn]{F.~Broggi}
\author[infn]{C.~Curatolo}
\author[lnf]{A.~Marocchino}
\author[unimi]{N.~Panzeri}
\author[infn,unimi]{V.~Petrillo}
\author[infn]{A.~Rossi}
\author[infn,unimi]{M.~Rossetti~Conti}

\address[infn]{INFN-Sezione di Milano, via Celoria 16, 20133 Milano, Italy}

\address[lnf]{Laboratori  Nazionali  di  Frascati,  Via  Enrico  Fermi  40,  00044  Frascati,  Italy\fnref{fn3}}

\address[unimi]{Universita degli Studi di Milano, via Celoria 16, 20133 Milano,
Italy\fnref{fn2}}

\begin{abstract}
The conceptual design of a compact source of GeV-class muons is presented, based on a plasma based electron-gamma collider. Evaluations of muon flux, spectra and brilliance are presented, carried out with ad-hoc montecarlo simulations of the electron-gamma collisions. These are analyzed in the context of a large spread of the invariant mass in the e-gamma interaction, due to the typical characteristics of plasma self-injected GeV electron beams, carrying large bunch charges with huge energy spread. The availability of a compact point-like muon source, triggerable at nsec level, may open a completely new scenario in the muon radiography application field.

\end{abstract}


\end{frontmatter}


\section{Introduction}
Muons presence on the Earth is due to the interaction between cosmic rays and the atmosphere: they are produced from pion $\pi^{\pm}$ and kaon $\textit{K}^{\pm}$ decay in the high atmosphere ($15000\;m$), and they reach our planet surface with a medium kinetic energy of $4\;GeV$. Atmospheric muons flux is spread all over the solid angle $\Omega$, and it's measured to be \cite{Bugaec,Morris} $\dfrac{dN}{d\Omega dt}=0.66\cos^2\left(\theta_z\right)\;{sr^{-1}.cm^{-2}.min^{-1}}$, i.e. integrating over the upper hemisphere solid angle and considering only muons with momentum $p_{\mu}>1\;GeV.c^{-1}$, $\dfrac{dN}{dAdt}\simeq 1\;cm^{-2}.min^{-1}$. Roughly, we could say that an human hand, anyway orientated, is crossed by a muon every minute.

GeV-class muons are keys to several strategic applications, in particular radiography of very thick objects (Volcanoes, Nuclear Power Plants, National Security) thanks to their high penetration/low stopping power (compared to photons/electrons\dots). A compact muon source based on the most advanced technologies could deliver a muon beam with $1-100\;muons/sec$ at GeV energy, collimated within hundreds/tens mrads. A Plasma Accelerator could guarantee the needed compactness of a GeV muon source (order of magnitude cheaper and shorter than GeV-class muon section of a typical muon collider). The combination of advancement in plasma accelerators (high charge GeV electron bunches) and in Compton Sources (high intensity tens
MeV-class photon beams as in ELI-NP-GBS) allows to conceive such a source possible in the near future. The challenge consists in running a $10^{31}$ $cm^{-2}s^{-1}$ luminosity (Lorentz Boosted) $e-\gamma$ collider at $E_{cm}=400$ $MeV$ to make a point-like, $GeV$-class, $nsec$ synchronized, muon source at $1\;\mu^{+,-}/s$ with collimated emission ($200\;mrad$) compact Muon Photo-Cathode producing $\mu$-pairs with $GeV$-scale energy .

The basic ingredients of a plasma based muon source are: a laser driven self-injected plasma accelerator generating low-quality (large energy spread, large emittance) high charge ($10\;nC$) electron bunch at $E > 1.5\; GeV$ (cfr. state of the art: $100\;pC$ at $5\;GeV$, $1\; nC$ at $500\; MeV$); a high-power interaction laser ELI-NP-GBS (Yb:Yag $1\; J\; @ \;1 \;kHz$, state of the art $1\; J\; @\; 100\; Hz$). Control, reproduce, stabilize the $e-\gamma$ collisions at IP with $\mu m$-size beam spots within the gas jet of plasma accelerator. Embed the whole accelerator ($3-5\;m$ in size) into a thick radio-protection bunker absorbing all beams but the muons (escaping through bunker walls). Additional filtering of surviving $e^-,\; \gamma's$ w.r.t. positive muons through magnetic fan-out spectrometer. A proof-of-principle experiment can be proposed: $0.1-1\; m$-pair per second gated in $10\;nsec$ time frame covering a $4\;m^2$ detector located $3\;m$ far from the point-like source (cfr. $400*10^{-8}*100 = 4*10^{-4}$ atmospheric muons \ensuremath{\Rightarrow} SNR $> 250$). 

\section{Muon photoproduction}
\begin{figure}
\centering
\includegraphics[width=0.7\linewidth]{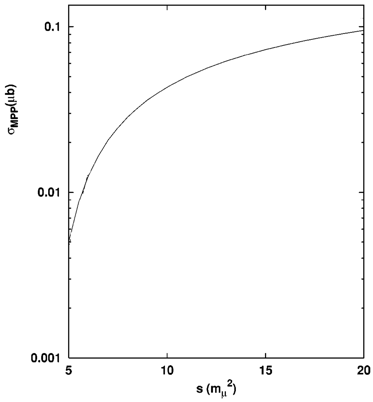}\caption{$\sigma_{MPP}$ as function of square of centre-of-mass energy $s$
(in unit of $m_{\mu}^{2}$) }
\label{fig01} 
\end{figure}

For the muon source we consider the process:

	\begin{equation}
	e^-+\gamma\rightarrow\mu^{+}+\mu^{-}+e \\
	\nonumber
	\end{equation}

where the muon flux scales with the total cross section \cite{Atar}:

	\begin{equation}
	\sigma_{MPP}(s)\simeq\frac{2\alpha^3}{m^2_{\mu}}\ln(2)\ln\left(\frac{s}{m^2_e}\right)
	\end{equation}

 as function of the invariant mass $s=E_{cm}^{2}/m_{\mu}^{2}$, where $m_{\mu,e}$ is muons and electrons rest mass, $\alpha$ is fine structure constant,  $E_{cm}=2\sqrt{\gamma m_{e}h\nu}$ is deduced from the kinematics of the system, where $h\nu$ is energy of laser photon. Since the cross section has a cut-off at $2m_{\mu}$, to exceed this energy cut-off we need to have an energy in the center of mass of the system bigger than about $200\;MeV$. The cross-section of $e+\gamma\rightarrow\mu^{+}+\mu^{-}+e$ scattering is reported in Fig.\ref{fig01}.

\begin{figure}
\centering
\includegraphics[width=1\linewidth]{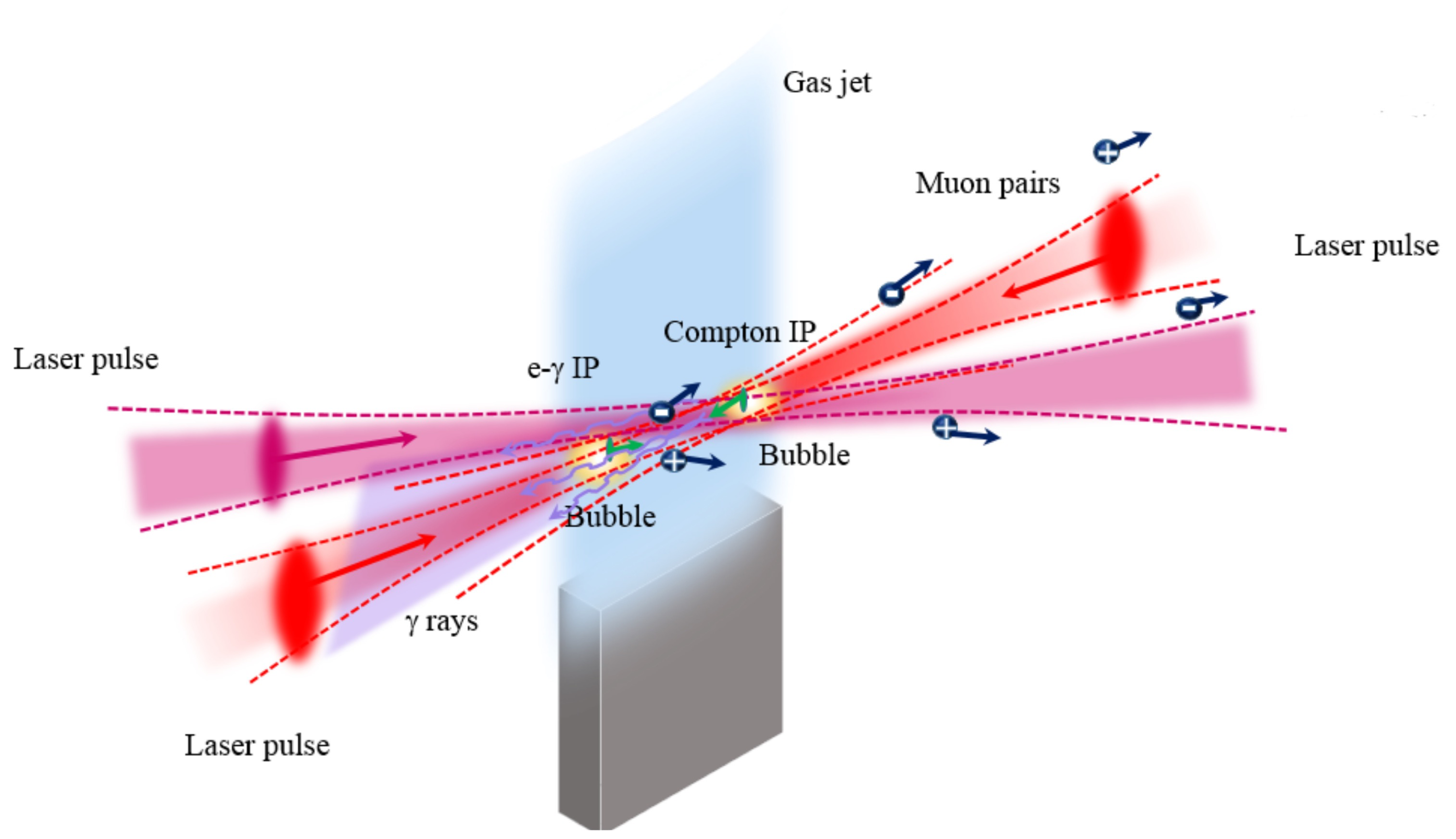}\caption{Sheme of source}
\label{fig02} 
\end{figure}

A possible implementation of the muon source relies on the interaction
between an electron beam accelerated by a plasma technique and a photon
beam and is shown in Fig. \ref{fig02}. 

If we consider an electron beam whose energy is of the order of
$1.5\;GeV$, the photons should gamma rays with tens $MeV$ of energy. Therefore
the gamma rays are foreseen to be produced by a Compton source that
can either use the same electron beam of the muon production, or another
beam produced ad-hoc for the Compton interaction. 

It is  important to notice that in this proposed scheme of $e^-,\gamma$ collision, the center of mass reference system moves relatively to the laboratory system with $\gamma_{cm}=\frac{E_{LAB}}{E_{CM}}\cong\frac{1}{2}\left(\sqrt{\frac{\gamma m_e}{h\nu}}\right)$, in the direction of the electron. Due to this fact, the muons created in center of mass system with energy close to rest mass energy will move with a $\gamma_{CM}$ in the laboratory system. As a consequence, the proposed scheme provides a beam of muon that has a $\gamma_{\mu}\geq\gamma_{CM}$.

We constructed an ideal electron beam, with parameters similar to
the state of the art of plasma accelerated beams \cite{chinesi,lifshits},
see Table \ref{tab01}.

\begin{table}
\centering
\caption{Electron beam parameters}
\begin{tabular}{|c|c|}
\hline 
Electrons energy $GeV$ & 1.6\tabularnewline
\hline 
Energy spread \% & 10\tabularnewline
\hline 
$\sigma_{x,y}\;\mu m$ & 1\tabularnewline
\hline 
$\sigma_{Px,Py}\;MeV$ & 5\tabularnewline
\hline 
$\sigma_{z}\;mm$ & 0.005\tabularnewline
\hline 
\end{tabular}
\label{tab01}
\end{table}

\begin{figure}
\centering
\includegraphics[width=1\linewidth]{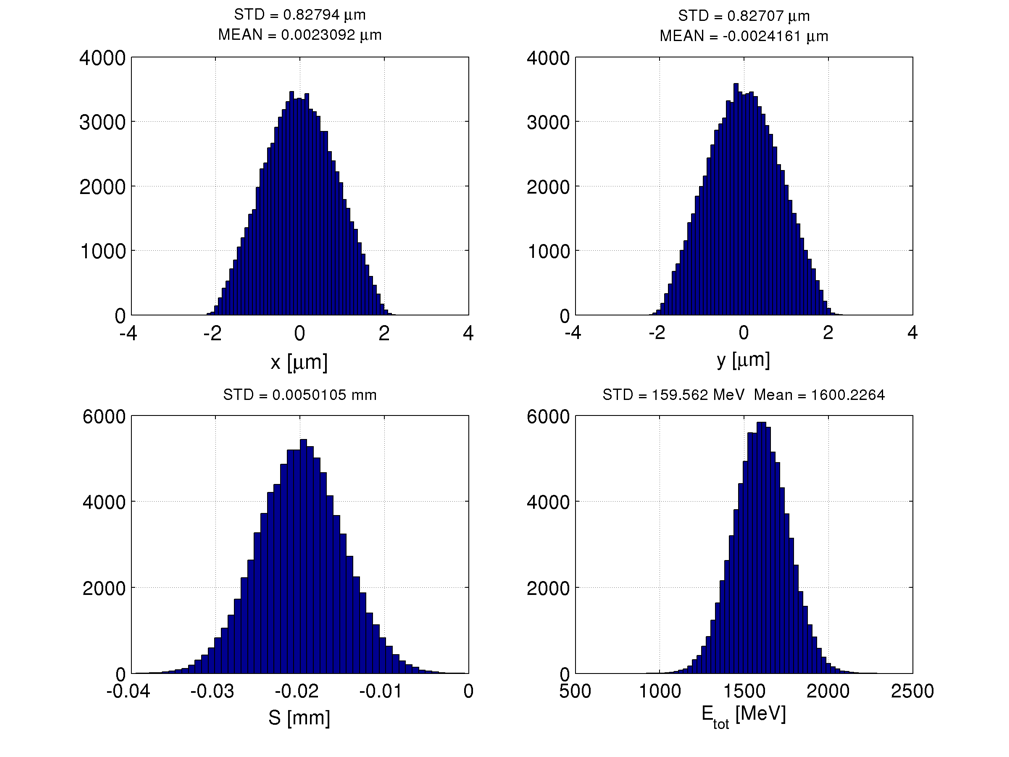}\caption{Electron beam phase-space}
\label{fig03} 
\end{figure}

 The phase spaces are in Fig. \ref{fig03}.
 
The Compton source exploits the same electron beam and a Yb:Yag
laser with parameters shown in Table \ref{tab02}.

\begin{table}
\centering
\begin{tabular}{|c|c|}
\hline 
Pulse energy $J$ & 1\tabularnewline
\hline 
wavelength $nm$ & 1030\tabularnewline
\hline 
$\sigma_{l}$$\mu m$ & 5\tabularnewline
\hline 
$\sigma_{t}$$ps$ & 1.5\tabularnewline
\hline 
\end{tabular}\caption{Laser parameters}
\label{tab02}
\end{table}

\begin{figure}
\centering
\includegraphics[width=1\linewidth]{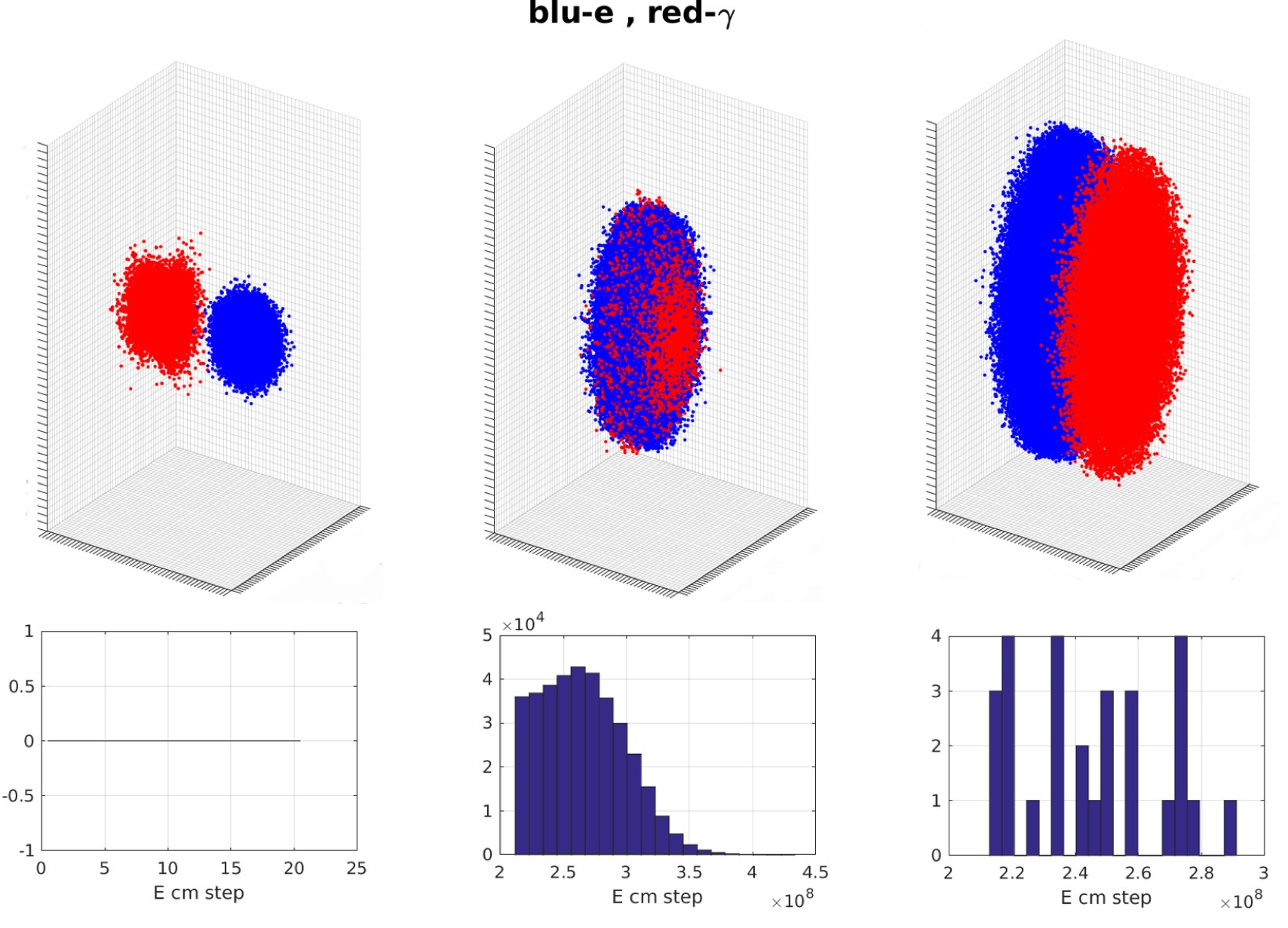}\caption{Propagation of particle in ROSE code at various instance. }
\label{fig04} 
\end{figure}

 The simulations of Compton $\gamma-$rays have been done with the
MonteCarlo code CAIN \cite{cain}. Fig. 4 shows the gamma spectrum,
with a total number of photon of about $1.7\cdot10^{11}$ and energy
$E_{ph}=4\gamma^{2}E$ laser ranging from $0$ up to $70\;MeV$. However
only photons with energy bigger than $25\;MeV$ partecipate to the muon
production. The propagation of the beams has been done with the code
ROSE \cite{ROSE}. The code ROSE (Rate Of Scattering Events) has been
implemented for studying the photon\textendash photon scattering
and then applied to other particle collisions and decays, as Breit\textendash Wheeler, TPP, Compton scattering \cite{micgg,drebgg,drebBW,drebBWipac}. Starting from two colliding beams of massive
particles or photons (say beam 1 and beam 2) defined through the
phase spaces of an appropriate number $N_{1;2}$ of macroparticles
of weight respectively $q_{1;2}$, the procedure requires the definition
of a common space grid where the kinematics takes place. The tracking
of both beams during their overlapping up to the end of the scattering
process permits to dimension the total space window. The initial time
$t_0$ is the instant when the first collisions occur, the time evolution
being discretized over a total of NT steps. Fig. \ref{fig04}

\begin{figure}
\centering
\includegraphics[width=0.9\linewidth]{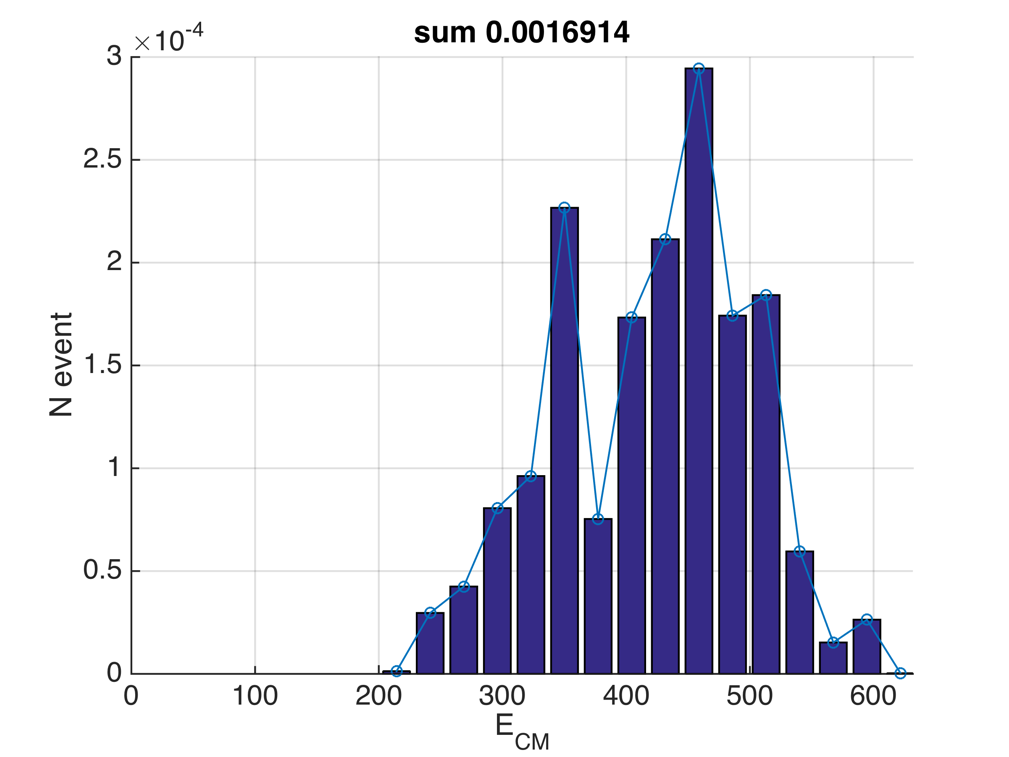}\caption{Energy distributions of muons}
\label{fig05} 
\end{figure}

 shows few temporal snapshot of the interaction (upper line) and the
number of interactions as function of the energy of the center of 
mass at the relative shot given by the convolution of the energy distribution
of the possible events with the cross section. The final output of
muons is presented in Fig.\;\ref{fig05}.

\begin{figure}
\centering
\includegraphics[width=0.9\linewidth]{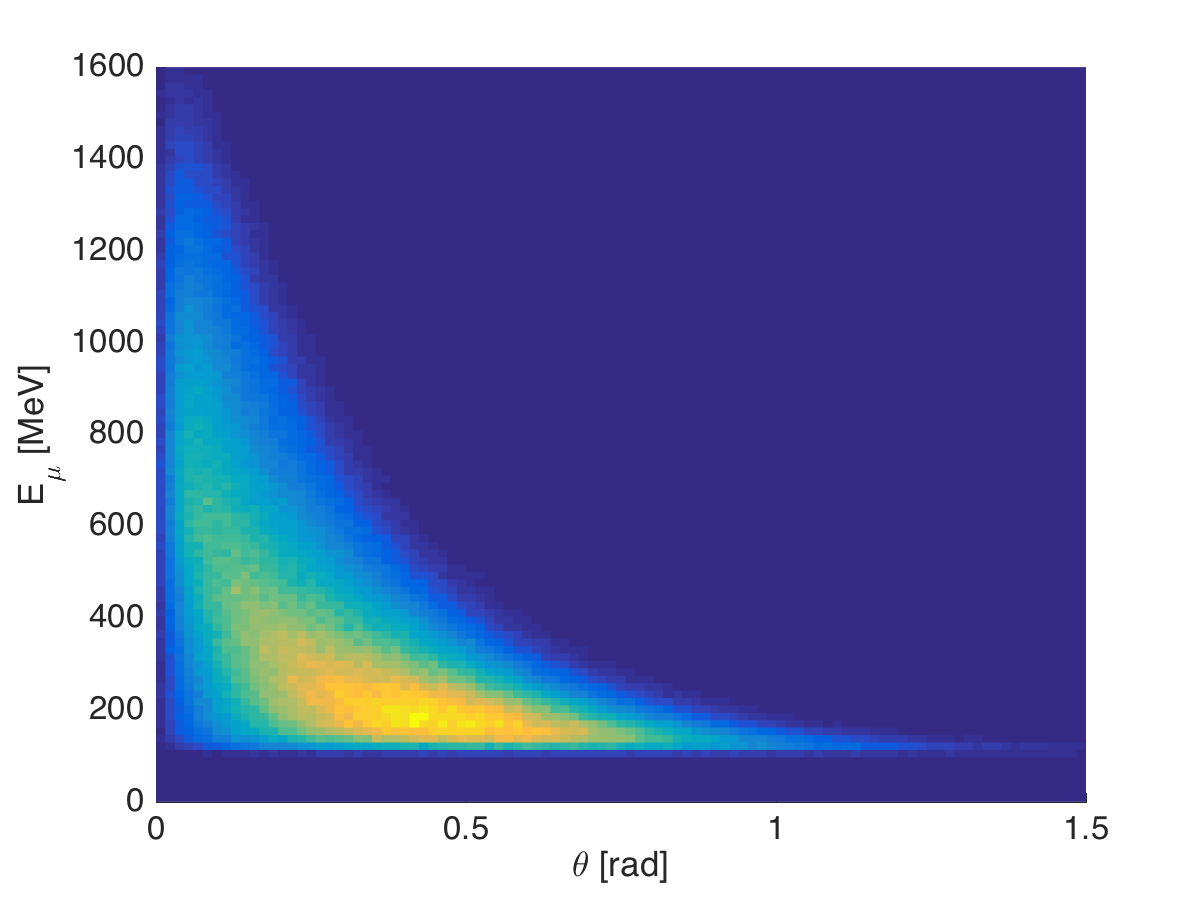}\caption{Energy-Angular distribution of muons}
\label{fig08} 
\end{figure}

To get the distributions of muons, simulations by event generator WHIZARD \cite{kilian} were done. The energy of the produced muons ranges between $105.65\;MeV$ (the muon is produced at rest, i.e. backward in CM) and about $2\;GeV$ (all of the electron energy is transferred to the muon). The energy distribution of the produced muons is peaked around $150\;MeV$. The almost totality of the muons is emitted in a cone of aperture $\theta$ of 1 radiant, the most part of them within a angle $1/\gamma_{cm}\simeq350$ milliradiants with the peak of emission around 100 millirads. The Energy-Angular distribution of muons presented in Fig.\;\ref{fig08}.

\begin{figure}
\centering
\includegraphics[width=1\linewidth]{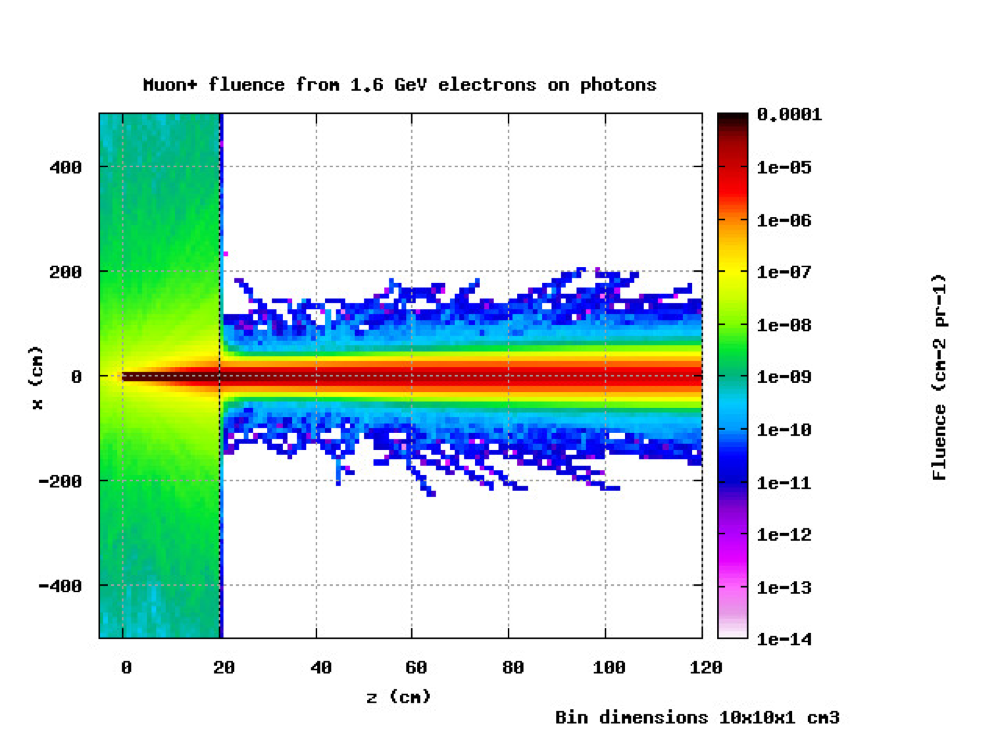}\caption{The number of emerging muon per primary incident muon for the case of $1\;m$ concrete wall.}
\label{fig09} 
\end{figure}

\begin{figure}
\centering
\includegraphics[width=1\linewidth]{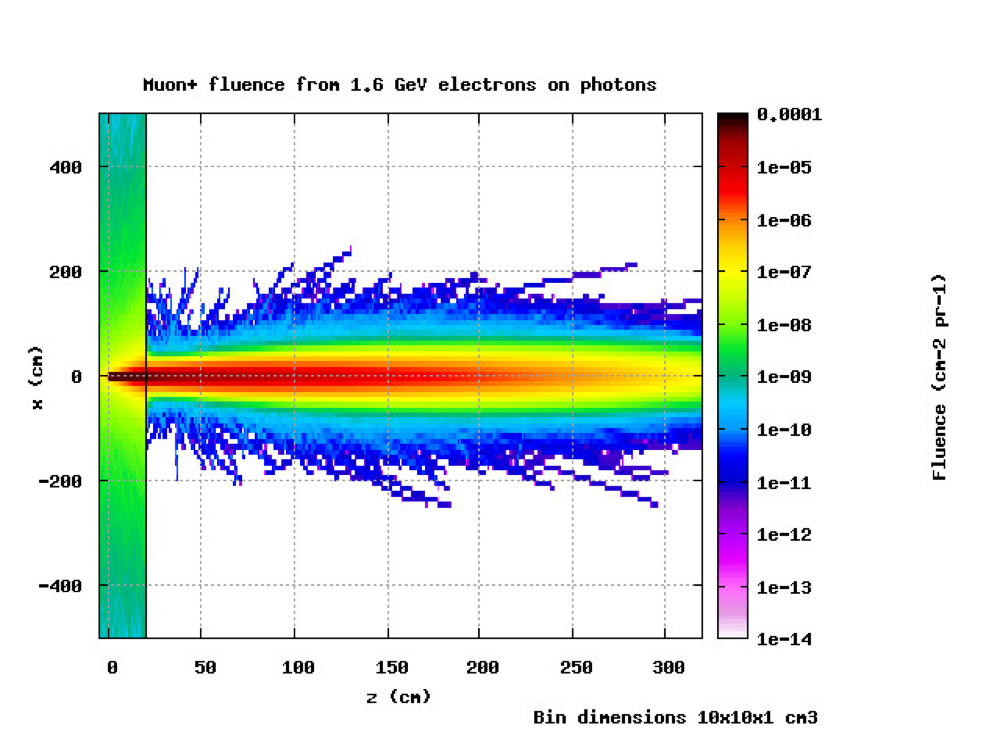}\caption{The number of emerging muon per primary incident muon for the case of $3\;m$ concrete wall.}
\label{fig10} 
\end{figure}

The produced muons are then driven to a concrete wall. The dynamics (or the interactions) of the muons inside the concrete wall has been evaluated by the FLUKA \cite{fasso,fluka} code
Two thickness have been considered $1\;m$ and $3\;m$ concrete. 
The results of the FLUKA evaluation, obtained with 20 different runs of the $999500$ primary muon are shown in the following plots. The cut off used in the simulations was $100\;keV$
The number of emerging muon per primary incident muon  is  $0.2442898 \pm 6.5563787E-03\%$ and $5.0838422E-03 \pm 0.1475075\%$ for the case of $1\;m$ and $3\;m$ concrete wall respectively presented on Fig \ref{fig09} and \ref{fig10}.

\begin{figure}
\centering
\includegraphics[width=0.9\linewidth]{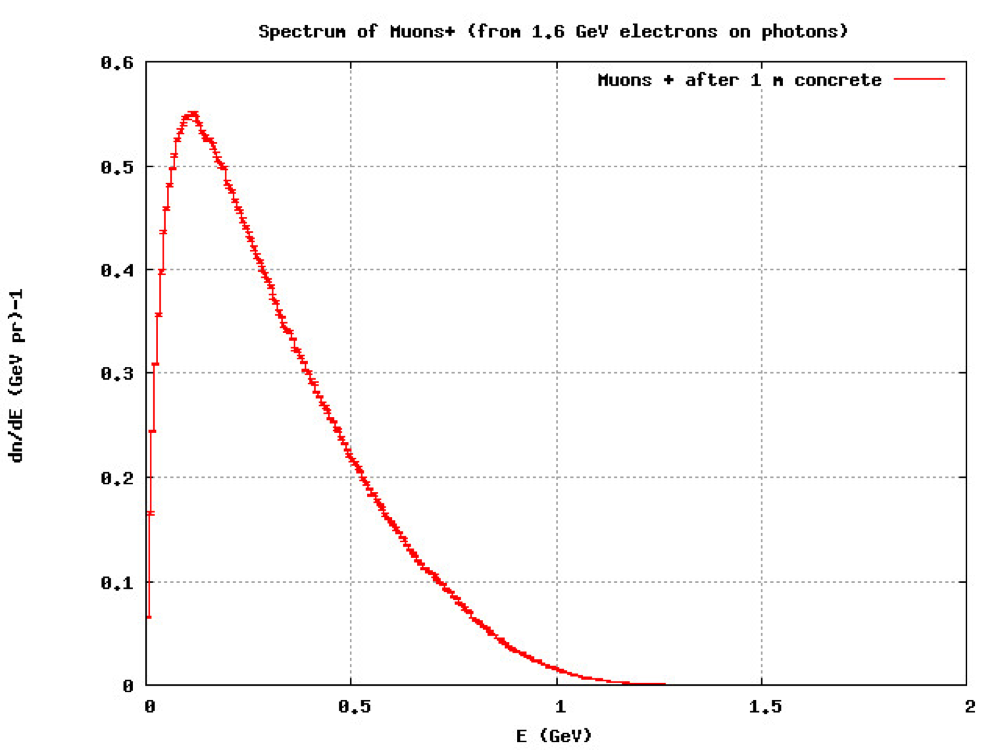}\caption{The spectrum of emerging muon per primary incident muon for the case of $1\;m$ concrete wall.}
\label{fig11} 
\end{figure}

\begin{figure}
\centering
\includegraphics[width=0.9\linewidth]{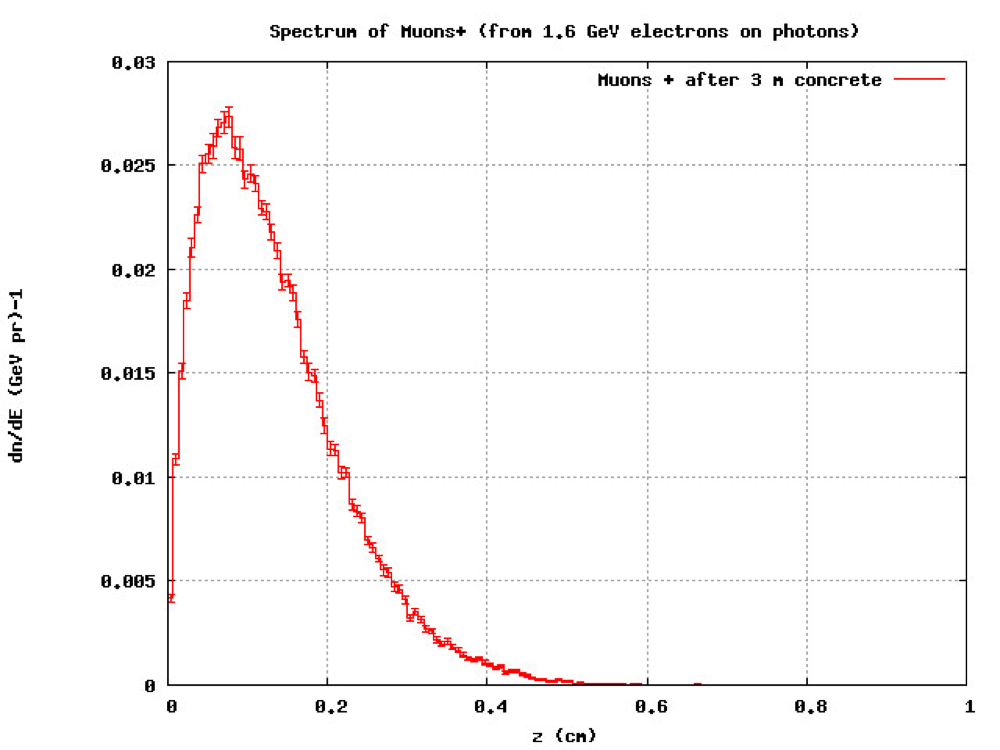}\caption{The spectrum of emerging muon per primary incident muon for the case of $3\;m$ concrete wall.}
\label{fig12} 
\end{figure}

The spectrum of the emitted muons (integrated  over the angle) is shown on the Fig. \ref{fig11} in case of $1\;m$ and on Fig. \ref{fig12} $3\;m$ concrete wall.

\begin{figure}
\centering
\includegraphics[width=1\linewidth]{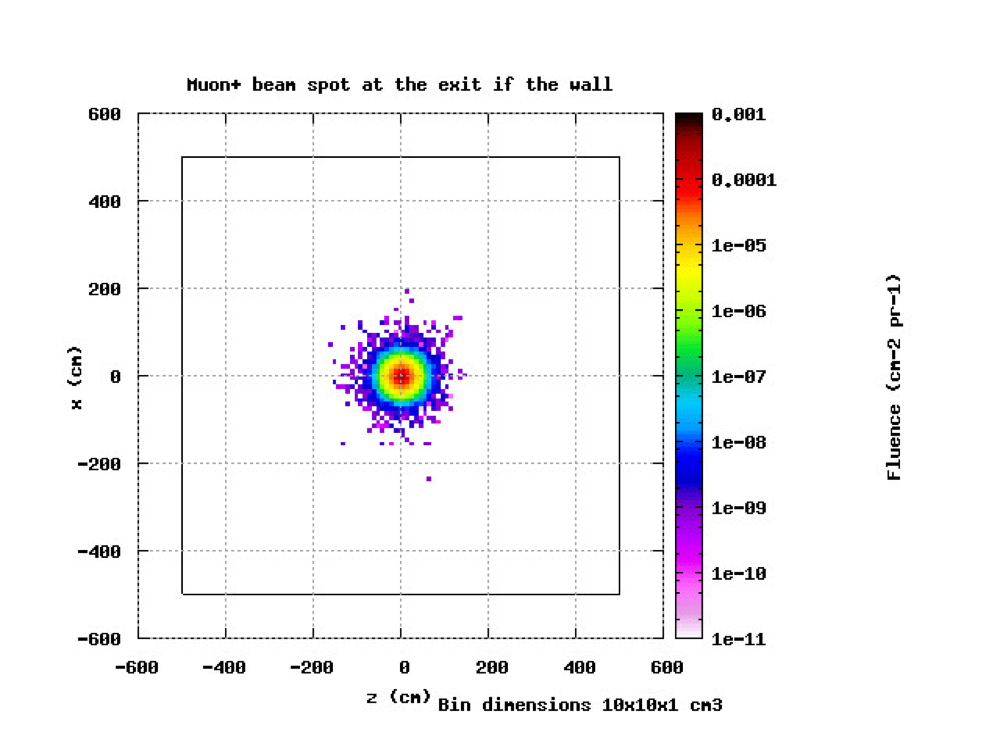}\caption{The beam spot of the muon beam at the exit of the concrete.}
\label{fig13} 
\end{figure}

The beam spot of the muon beam at the exit of the concrete is shown on Fig. \ref{fig13} .

\section{Conclusion}

Advancement in fiber-lasers, expected to meet laser-plasma based TeV
collider requirements at $10-100\;kHz$ rep rate, offers the opportunity to develop a Compact Muon Source delivering the muon beam with $1-100\;muons/sec$ at GeV energy, collimated within hundreds/tens mrads, synchronized at nsec
level, based on a compact O($10\;m$) and cheap O(10 M\euro) system. 

\label{}





\end{document}